\documentclass[11pt]{article}
\usepackage{latexsym}
\usepackage{array}
\usepackage{amsmath,amssymb,amsthm,amsfonts}
\usepackage{vmargin}
\usepackage{fancyhdr}

\setpapersize[portrait]{USletter}
\setmarginsrb{1in}{1in}{1in}{1in}{0in}{0in}{12pt}{24pt}
\newcommand{\thmabove}{7.5pt}
\newcommand{\thmbelow}{0pt}
\newcommand{\proofbelow}{0pt}
\renewcommand{\thmbelow}{6pt}
\renewcommand{\proofbelow}{7pt}
\makeatletter
\renewcommand{\maketitle}{
    \begin{center}
    \begin{minipage}[t]{6.5in}
    \begin{center}
    \vspace*{15pt}
    {\@title \par}
    \vspace*{20pt}
    {\@author}
    \vspace*{3pt}
    \end{center}
    \end{minipage}
    \end{center}
}
\makeatother
\renewcommand{\epsilon}{\varepsilon}
\renewcommand{\ldots}{...}

\renewcommand{\th}{\ifmmode{^{\textrm{th}}}\else{\textsuperscript{th}\ }\fi}

\newcommand{\plusminus}{\pm}
\newcommand{\abs}[1]{\lvert #1 \rvert}
\newcommand{\union}{\cup}

\newcommand{\intersect}{\cap}                                   

\newcommand{\set}[1]{\left \{ #1 \right \}}                     
        
\newcommand{\card}[1]{\abs{#1}}
\newcommand{\sumstack}[1]{\sum_{\substack{#1}}}

\newcommand{\ceil}[1]{\left\lceil #1 \right\rceil}

\newcommand{\smallfrac}[2]{{\textstyle \frac{#1}{#2}}}
\newcommand{\smallbinom}[2]{{\textstyle \binom{#1}{#2}}}

\newcommand{\transpose}{^{\textsf{T}}}                                    

\renewcommand{\det}{\operatorname{det}}
\newcommand{\rank}{\operatorname{rank}}

\newcommand{\minor}[3]{#1_{\mathrm{del}(#2,#3)}}

\newenvironment{smallpmatrix}                          
    {\left(\begin{smallmatrix}}
    {\end{smallmatrix}\right)}

\newcommand{\mat}{\mathbf{M}}
\newcommand{\freemat}{{\mathbf{F}}}
\newcommand{\dirsum}{\oplus}

\newcommand{\myhline}{\mbox{\rule[2.5pt]{14pt}{0.4pt}}\hspace*{2pt}}

\newcommand{\tC}{\tilde{C}}

\newcommand{\tM}{\tilde{M}}
\newcommand{\tN}{\tilde{N}}
\newcommand{\tO}{\tilde{O}}

\newcommand{\tQ}{\tilde{Q}}

\newcommand{\tU}{\tilde{U}}
\newcommand{\tV}{\tilde{V}}

\newcommand{\tZ}{\tilde{Z}}

\newcommand{\bF}{\mathbb{F}}

\newcommand{\cB}{\mathcal{B}}

\newcommand{\cI}{\mathcal{I}}

\newcommand{\cY}{\mathcal{Y}}

\newcommand{\EquationName}[1]{\label{eq:#1}}

\newcommand{\FactName}[1]{\label{fact:#1}}

\newcommand{\LemmaName}[1]{\label{lem:#1}}

\newcommand{\TheoremName}[1]{\label{thm:#1}}
\newcommand{\SectionName}[1]{\label{sec:#1}}

\newcommand{\Equation}[1]{Eq.\:\eqref{eq:#1}}

\newcommand{\Fact}[1]{Fact~\ref{fact:#1}}

\newcommand{\Lemma}[1]{Lemma~\ref{lem:#1}}

\newcommand{\Section}[1]{Section~\ref{sec:#1}}

\newcommand{\Theorem}[1]{Theorem~\ref{thm:#1}}

\newtheoremstyle{mythmstyle}
  {\thmabove}   
  {\thmbelow}   
  {}            
  {}            
  {\bfseries}   
  {. }          
  {2.5pt}       
  {\thmname{#1}\thmnumber{ #2}\thmnote{ \normalfont (#3)}}   
\theoremstyle{mythmstyle}
\newtheorem{theorem}{Theorem}
\newtheorem{corollary}[theorem]{Corollary}

\newtheorem{fact}[theorem]{Fact}
\newtheorem{lemma}[theorem]{Lemma}

\newtheorem{myexample}[theorem]{Example}
\newcommand{\afterproof}{\hfill $\blacksquare$ \par \vspace{\proofbelow}}
\renewenvironment{proof}{\noindent\textbf{Proof}.\,}{\afterproof}

\allowdisplaybreaks[3]

\newcommand{\defeq}{:=}
\newcommand{\newterm}[1]{\textit{#1}}

\newcounter{itemctr}
\renewenvironment{enumerate}{
    \begin{list}{\textrm{(\arabic{itemctr}):}}{
        \usecounter{itemctr}\setlength{\labelsep}{5mm}\setlength{\leftmargin}{15mm}
        \setlength{\labelwidth}{15mm}\setlength{\itemsep}{0mm}\setlength{\parsep}{1mm}
        \setlength{\topsep}{0mm}\setlength{\listparindent}{0pt}
    }
}
{
    \end{list}
}

\newlength\abovesectionskip
\newlength\belowsectionskip
\def\sectionfont{\normalfont\Large\bfseries}

\newlength\abovesubsectionskip
\newlength\belowsubsectionskip
\def\subsectionfont{\normalfont\large\bfseries}

\newlength\abovesubsubsectionskip
\newlength\belowsubsubsectionskip
\def\subsubsectionfont{\normalfont\bfseries}
\makeatletter
\def\section{\@startsection{section}{1}{\z@}{-\abovesectionskip}%
               {\belowsectionskip}{\sectionfont}}
\def\subsection{\@startsection{subsection}{2}{\z@}{-\abovesubsectionskip}%
                  {\belowsubsectionskip}{\subsectionfont}}
\def\subsubsection{\@startsection{subsubsection}{2}{\z@}{-\abovesubsubsectionskip}%
                  {\belowsubsubsectionskip}{\subsubsectionfont}}
\makeatother

\title{
{\Large\bf Algebraic Structures and Algorithms \\
for Matching and Matroid Problems} \\
{\Large (Preliminary Version)}}
\author{
\large {Nicholas J.\ A.\ Harvey\/}%
\footnotemark[1]
\\
MIT Computer Science and Artificial Intelligence Laboratory \\
\texttt{nickh@mit.edu}
}

\begin{document}
\maketitle
\footnotetext[1]{Supported by a Natural Sciences and Engineering Research Council of Canada PGS Scholarship.}

\begin{abstract}
Basic path-matchings, introduced by Cunningham and Geelen (FOCS 1996),
are a common generalization of matroid intersection and non-bipartite matching.
The main results of this paper are a new algebraic characterization of basic path-matching problems
and an algorithm for constructing basic path-matchings in $\tO(n^\omega)$ time,
where $n$ is the number of vertices and $\omega$ is the exponent for matrix multiplication.
Our algorithms are randomized, and our approach assumes that the given matroids are linear and can
be represented over the same field.

Our main results have interesting consequences for several special cases of path-matching problems.
For matroid intersection, we obtain an algorithm with
running time $\tO(nr^{\omega-1})=O(nr^{1.38})$,
where the matroids have $n$ elements and rank $r$.
This improves the long-standing bound of $O(nr^{1.62})$ due to Gabow and Xu (FOCS 1989).
Also, we obtain a simple, purely algebraic algorithm for non-bipartite matching with running time
$\tO(n^\omega)$.
This resolves the central open problem of Mucha and Sankowski (FOCS 2004).
\end{abstract}

%%%%%%%%%%%%%%%%%%%%%%%%%%%%%%%%%%%%%%%%%%%%%%%%%%%%%%%%%%%%%%%%%%%%%%%%%%%%%%%%%%%%%%%%%%%%%%%%%%%%%
%%%%%%%%%%%%%%%%%%%%%%%%%%%%%%%%%%%%%%%%%%%%%%%%%%%%%%%%%%%%%%%%%%%%%%%%%%%%%%%%%%%%%%%%%%%%%%%%%%%%%
%%%%%%%%%%%%%%%%%%%%%%%%%%%%%%%%%%%%%%%%%%%%%%%%%%%%%%%%%%%%%%%%%%%%%%%%%%%%%%%%%%%%%%%%%%%%%%%%%%%%%

\section{Introduction}

Non-bipartite matching is a fundamental problem that has played a pivotal
role in the development of graph theory and combinatorial optimization \cite{LovaszPlummer}.
Matroid intersection is another fundamental problem; its min-max characterization
and efficient algorithms have had far-reaching consequences in both combinatorics and
computer science \cite{Schrijver}.
Basic path-matchings, due to Cunningham and Geelen \cite{CunninghamGeelenFOCS,CunninghamGeelen},
are an elegant generalization of these two classic problems.

Cunningham and Geelen showed that maximum-weight basic path-matchings can be found in
polynomial time via the ellipsoid method \cite{CunninghamGeelen}.
Later, path-matchings were generalized to even-factors
in a class of digraphs known as ``weakly-symmetric''.
Cunningham and Geelen \cite{CunninghamGeelenManuscript} devised
a combinatorial algorithm to compute a maximum even-factor in such graphs.
Combining this with a matroid (or valuated matroid \cite{MurotaValuatedMatroid}) intersection algorithm,
they obtain an algorithm to compute a maximum (or maximum-weight) basic even-factor.
Pap \cite{Pap} considers even-factors in a class of digraphs known as ``odd-cycle symmetric'',
and gives a combinatorial algorithm to find a maximum even-factor in such graphs.
Finally, Takazawa \cite{Takazawa} gives a strongly-polynomial algorithm to compute
maximum-weight even-factors in odd-cycle symmetric graphs.

Matroid intersection algorithms have a more storied history.
Polynomial time algorithms for matroid intersection were developed in the 1970s
by various authors \cite{EdmondsSubmodular,EdmondsIntersection,LawlerAlg}.
The efficiency of these early algorithms was typically measured relative to an oracle for testing
independence.
Cunningham~\cite{Cunningham} later developed an efficient algorithm for intersection of
linear matroids (those that can be represented as a matrix),
which is the broadest class of matroids with efficient representations.
Cunningham's algorithm requires $O(nr^2 \log r)$ time,
where $n$ denotes the size of the ground set and $r$ bounds the rank of the matroids.
Gabow and Xu \cite{GabowXuFocs,GabowXu} obtained an improved bound of
$O(nr^{1.62})$ through the use of fast matrix multiplication and quite technical arguments.
However, their bound does not seem to be a natural one;
Gabow and Xu explicitly pose the problem of improving their bound as an open question.

Several variants and generalizations of matroid intersection have
been considered, notably matroid matching. Surprisingly, matroid
matching problems on general graphs are not solvable in polynomial
time (in the oracle model) \cite{LovaszPlummer}. On the other
hand, bipartite matroid matching problems are much more tractable,
and have close connections to matroid intersection
\cite{Fujishige}. It was shown by Edmonds \cite[Theorem
81]{EdmondsSubmodular} that bipartite matroid matching reduces to
matroid intersection, and therefore can be solved efficiently.
Efficient algorithms for bipartite matroid matching were implicit
in Aigner-Dowling \cite{Aigner} and explicit in Tomizawa-Iri
\cite{Tomizawa}.

The aforementioned algorithms for matroid intersection and bipartite matroid matching
are all based on augmenting paths;
indeed, so are most algorithms for these problems
\cite{Aigner,Cunningham,EdmondsIntersection,Frank,Fujishige,GabowXuFocs,GabowXu,LawlerAlg,Tomizawa}.
In contrast, Lov\'asz \cite{RandomMatching} introduced a randomized algebraic approach for matroid
matchings; this approach was employed and extended by Barvinok \cite{Barvinok}
and Camerini et al.~\cite{Camerini}.
This paper, building on results of Geelen \cite{GeelenThesis},
extends the algebraic framework to basic path-matching problems.

Algebraic approaches have recently been used by Mucha and Sankowski
to obtain efficient algorithms for matching problems in bipartite and general graphs \cite{Mucha}.
For bipartite graphs, they consider the Edmonds matrix and apply a simple but elegant variant
of the Hopcroft-Bunch Gaussian elimination algorithm \cite{Bunch}.
For general graphs, their algorithm is much more complicated:
it maintains the canonical partition of the graph using sophisticated data structures
for testing dynamic connectivity.
Sankowski \cite{SankowskiParallel} also developed an $\text{RNC}^5$ algorithm for
constructing perfect matchings that uses only $\tO(n^\omega)$ processors,
yielding another sequential algorithm that uses only $\tO(n^\omega)$ time.
However, this algorithm is based on sophisticated parallel subroutines for
evaluating a polynomial at a matrix and for computing the characteristic polynomial of a matrix.

%%%%%%%%%%%%%%%%%%%%%%%%%%%%%%%%%%%%%%%%%%%%%%%%%%%%%%%%%%%%%%%%%%%%%%%%%%%%%%%%%%%%%%%%%%%%%%%%%%%%%
\subsection{Our results}

The main results of this paper are a new algebraic characterization of basic path-matching problems
and a randomized algorithm for constructing basic path-matchings in $\tO(n^\omega)$ time,
where $n$ is the number of vertices and $\omega<2.38$ is the exponent for matrix multiplication
\cite{Coppersmith}.
Our approach involves two assumptions.
First, as is common, we assume that the matroids associated with the
basic path-matching problem are linear.
Additionally, we make the mild technical assumption that any given pair of
matroids are represented as matrices over the same field.
Although there exist matroids for which this assumption
does not hold (e.g., the Fano and non-Fano matroids),
this assumption can be satisfied for the vast majority of matroids arising in applications.
The regular matroids are those that are representable over all fields;
this class includes the graphic, cographic and partition matroids.
Many classes of matroids are representable over all but finitely many fields;
these include the uniform, matching, and transversal matroids,
as well as deltoids and gammoids \cite{Schrijver}.
Our results apply to any matroids from these classes.

As a consequence, we obtain $\tO(n^\omega)$ randomized algorithms
for constructing perfect path-matchings, maximum matchings in general graphs,
maximum bipartite matroid matchings, and maximum intersections of two matroids.
Here $n$ denotes the number of vertices in the graphs or, for matroid problems,
the cardinality of the ground set.
Our algorithms are purely algebraic and do not require use of
sophisticated data structures or subroutines other than matrix multiplication.
Therefore our non-bipartite matching algorithm resolves the central open question from the paper of
Mucha and Sankowski \cite{Mucha}.
Their structural approach using the canonical partition is not necessary;
a simple, but subtle, divide-and-conquer approach suffices.

For matroid problems, we can obtain tighter bounds by also considering the
rank $r$ of the given matroids.
For bipartite matroid matching, the bound improves only slightly:
the time required is $\tO(n^2 r^{\omega-2} + r^{\omega + 1})$,
which is tighter than $\tO(n^\omega)$ when $r = o(n^{\omega/(\omega+1)})$.
For matroid intersection problems, the improvement is considerable:
we present an algorithm that requires only $\tO(nr^{\omega-1})$ time.
This is essentially the best that one could hope for
since computing the rank of a $n \times r$ matrix reduces straightforwardly to
linear matroid intersection, and $O(nr^{\omega-1})$ is the best known bound for
computing the rank.
Interestingly, our matroid intersection algorithm resembles a well-known proof of the matroid
intersection theorem, for which it is remarked that the ``proof gives no hint of how to find the
[matroid intersection] efficiently'' \cite[p289]{Cook}.
Now suppose that we consider only algorithms using naive matrix multiplication.
Under this restriction, the best known algorithm for linear matroid intersection is
Cunningham's, which uses $O(nr^2 \log r)$ time.
The Gabow-Xu algorithm, despite its much more sophisticated techniques,
also achieves the same bound.
Our algorithm requires only $O(nr^2)$ time, under the assumption that the field over which the
matroids are represented is sufficiently large.

%%%%%%%%%%%%%%%%%%%%%%%%%%%%%%%%%%%%%%%%%%%%%%%%%%%%%%%%%%%%%%%%%%%%%%%%%%%%%%%%%%%%%%%%%%%%%%%%%%%%%
\subsection{Notation and Basic Facts}

The set of integers $\set{1,\ldots,m}$ is denoted $[m]$.
If $J$ is a set, $J+i$ denotes $J \union \set{i}$.
The $i\th$ elementary vector is denoted $e_i$,
that is, $e_i$ is $1$ in the $i\th$ component and zero elsewhere.
If $M$ is a matrix,
a submatrix containing rows $S$ and columns $T$ is denoted $M[S,T]$.
A submatrix containing all rows (columns) is denoted $M[*,T]$ ($M[S,*]$).
A submatrix $M[S,T]$ is sometimes written as $M_{S,T}$ when this enhances legibility.
The $i\th$ row (column) of $M$ is denoted $M_{i,*}$ ($M_{*,i}$).
An entry of $M$ is denoted $M_{i,j}$.
A submatrix containing rows $\set{a,\ldots,b}$ and columns $\set{c,\ldots,d}$ is denoted
$M_{a:b,\:c:d}$.
The submatrix obtained by deleting row $i$ and column $j$ (row-set $I$ and column-set $J$)
from $M$ is denoted $\minor{M}{i}{j}$ ($\minor{M}{I}{J}$).
If $u$ and $v$ are vectors and $c$ is a scalar,
the matrix $\tM=M + c u v\transpose$ is called a rank-1 update of $M$.
Assume that $M$ is non-singular and let $\alpha = c^{-1} + v \transpose M^{-1} u$.
The inverse of $\tM$ exists iff $\alpha \neq 0$, and equals
\begin{equation}
\EquationName{rank1}
\tM^{-1} \:=\:
M^{-1} - \alpha^{-1}
\big( M^{-1} \, u \big) \, \big( v\transpose \, M^{-1} \big),
\end{equation}
which is itself a rank-1 update of $M^{-1}$.
The following expansion of the determinant is useful.

\begin{fact}[Generalized Laplace expansion]
\FactName{laplace}
Let $M$ be an $n \times n$ matrix.
Fix a set of rows $I$ such that $\emptyset \neq I \subset \set{1, \ldots, n}$.
Then
\begin{equation}
\EquationName{laplace}
\det{M} ~=~ \sumstack{J \subset \set{1, \ldots, n} ,\, \card{J}=\card{I}}
\det{M[I,J]} \cdot \det{M[\bar{I},\bar{J}]} \cdot
(-1)^{\sum_{i \in I} i + \sum_{j \in J} j}.
\end{equation}
An analogous statement holds for a set of rows $I$ by taking the transpose of $A$.
\end{fact}
\begin{proof}
See Aitken \cite{Aitken} or Murota \cite{Murota}.
\end{proof}

A matroid is a tuple $\mat=(V,\cI,\cB,r)$ where $V$ is the ground set,
$\cI \subseteq 2^V$ is the collection of independent sets,
$\cB \subseteq \cI$ is the collection of bases,
and $r$ is the rank function.
The rank of the matroid $\mat$ is $r(V)$.
Together with $V$, any one of $\cI$, $\cB$, and $r$ is sufficient to specify the matroid,
so we do not necessarily mention all of them.
To emphasize connection to a specific matroid, we sometimes use the notation
$\cI_{\mat}$, $\cB_{\mat}$ and $r_{\mat}$.
For $J \subseteq V$, $\mat/J$ denotes the matroid obtained from $\mat$ by contracting $J$.
Its rank function is $r_{\mat/J}(S) \defeq r_\mat(J \union S) - r_\mat(J)$.
For $J \subseteq V$, $\mat \setminus J$ denotes the deletion of $J$ from $\mat$
(or restriction of $\mat$ to $V \setminus J$).
The direct sum of two matroids $M_1$ and $M_2$ on disjoint ground sets is denoted $M_1 \dirsum M_2$.
The free matroid on a set $V$ is $\freemat(V) \defeq (V,\cB)$ where $\cB=\set{V}$.

%%%%%%%%%%%%%%%%%%%%%%%%%%%%%%%%%%%%%%%%%%%%%%%%%%%%%%%%%%%%%%%%%%%%%%%%%%%%%%%%%%%%%%%%%%%%%%%%%%%%%
\section{Path-Matchings}
\SectionName{pathmatching}

\subsection{Definitions}

An instance of the a basic path-matching is a tuple $G=(T_1,T_2,S,E,\mat_1,\mat_2)$
where $(T_1 \union T_2 \union S, E)$ is a graph and each $\mat_i$ is a matroid $(T_i, \cI_i, r_i)$.
The vertex sets $T_1$, $T_2$ and $S$ are disjoint and furthermore $T_1$ and $T_2$ are stable sets
(no edge has both endpoints in either $T_1$ or $T_2$).
Let $s = \card{S}$ and assume that $\card{T_1}=\card{T_2}=t$,
and that $\mat_1$ and $\mat_2$ have the same rank $r$.

A \newterm{path-matching} is a collection of node-disjoint paths
with one endpoint in $T_1$ and the other in $T_2$,
together with a matching on the $S$-vertices not contained in any of the paths.
If $M \subseteq E$ is a path-matching, let $\partial_i M \subseteq T_i$ denote the set of vertices
in $T_i$ that are covered by $M$,
and let $\partial_S M \subseteq S$ denote the covered $S$-vertices.
A \newterm{perfect path-matching} is a path-matching $M$
such that $\partial_i M = T_i$ and $\partial_S M = S$.
A \newterm{basic path-matching} (bpm) is a path-matching
such that $\partial_i M \in \cB_{\mat_i}$ and that $\partial_S M = S$.

\subsection{Contracted Instances}
\SectionName{contracted}

Let $G=(T_1,T_2,S,E,\mat_1,\mat_2)$ be an instance of the basic path-matching problem.
A set $M \subseteq E$ is called an \newterm{extensible set} for $G$ if there exists a bpm $M' \supseteq M$.
Let $M$ be an extensible set and note that $\partial_i M \in \cI_{\mat_i}$.
We will now define a basic path-matching problem $G(M)$, which we call the \newterm{contraction}
of $G$ by $M$.

Define $\overline{\partial_i M} = T_i \setminus \partial_i M$,
and note that $\partial_1 M$ and $\partial_2 M$ are not necessarily equicardinal.
Let $P_i$ be the set of paths in $M$ with one endpoint in $T_i$ and the other in $S$.
The set $C_i \subseteq S$ consists of the endpoints in $S$ of the paths in $P_i$.
Define $T_i' \defeq \overline{\partial_i M} \union C_i$ and $S' \defeq S \setminus \partial_S M$.
Define $E' \subseteq E$ to be the set of edges with both endpoints in $S'$
or with one endpoint in $T_i'$ and the other in $S' \union T_j'$ where $i \neq j$.
The matroid $\mat_i'$ is $(\mat_i / \partial_i M) \oplus \freemat(C_i)$.
The contraction of $G$ by $M$ is $G(M) \defeq (T_1', T_2', S', E', \mat_1', \mat_2') $.

We now observe certain properties of $G(M)$.
First, $C_i \intersect S' = \emptyset$ since $C_i \subseteq \partial_S M$.
Second, the ground set of $\mat_i'$ is $T_i'$ and furthermore $r_{\mat_i'}(T_i') = r$.
Now let $p$ be the number of paths in $M$ connecting $T_1$ to $T_2$.
Then $\card{T_i'} = \card{T_i} - p$ since $\card{\partial_1 M} = \card{C_1} + p$.
Next, if $M' \supseteq M$ is a bpm for $G$ then $M'' \defeq M' \setminus M$ is a bpm for $G(M)$.
Conversely, if $M''$ is a bpm of $G(M)$ then $M \union M''$ is a bpm of $G$.
This follows because every basis of $\mat_i'$ clearly contains $C_i$,
and hence every bpm of $G(M)$ covers $C_i$.

\subsection{Algebraic Structure}

Suppose that each $\mat_i$ is a linear matroid representable over a common field $\bF$.
Let $Q_1$ be an $r \times t$ matrix whose columns represent $\mat_1$ over $\bF$
and let $Q_2$ be a $t \times r$ matrix whose rows represent $\mat_2$ over $\bF$.
For notational convenience, we will let $Q_1^J$ denote $Q_1[*,J]$ and $Q_2^J$ denote $Q_2[J,*]$.
We now define a matrix of indeterminates $X$ which describes the graph underlying $G$.
The rows of $X$ are indexed by $T_1 \union S$ and the columns are indexed by $T_2 \union S$.
The entries are defined as $X_{i,j} = \plusminus x_{\set{i,j}}$,
where the signs are chosen such that $X[S,S]$ is skew-symmetric.

\begin{lemma}[Geelen \cite{GeelenThesis}]
\LemmaName{geelen}
$X[S,S]$ is non-singular iff $G$ has a perfect path-matching.
\end{lemma}

We now extend this algebraic characterization to basic path-matching problems.
Let $D_i$ be a diagonal matrix of size $\card{\overline{\partial_i M}}$ whose entries contain
distinct indeterminates.
Define $Z(M)$ to be the following matrix:

$$ Z(M) = \begin{pmatrix}
                              & Q_1^{\partial_1 M} & Q_1^{\overline{\partial_1 M}} & \\
Q_2^{\partial_2 M}            &                    & \\
Q_2^{\overline{\partial_2 M}} &                    &                               & D_2 \\
                              &                    & D_1                           & X[\overline{\partial_1 M},\overline{\partial_2 M}] & X[\overline{\partial_1 M},C_2 \union S'] \\
                              &                    &                               & X[C_1 \union S',\overline{\partial_2 M}] & X[C_1 \union S',C_2 \union S']
\end{pmatrix},
$$
and for convenience let
$$ X' = \begin{pmatrix}
     & D_2 \\
 D_1 & X[\overline{\partial_1 M},\overline{\partial_2 M}] & X[\overline{\partial_1 M},C_2 \union S'] \\
     & X[C_1 \union S',\overline{\partial_2 M}] & X[C_1 \union S',C_2 \union S']
\end{pmatrix}.
$$

\vspace{\parsep}

\begin{theorem}
\TheoremName{matchingrank}
Let $M$ be a set of edges in $G$ such that $\partial_1 M \in \cI_{\mat_1}$
and $\partial_2 M \in \cI_{\mat_2}$.
Then $G(M)$ has a bpm iff $Z(M)$ is non-singular.
\end{theorem}

\begin{proof}
We apply the generalized Laplace expansion (\Fact{laplace}) to $Z(M)$ with the set of rows
$I = \set{1, \ldots, r}$, obtaining
\begin{equation}
\EquationName{expansion1}
\det Z(M) ~=~ \sumstack{A \subset [k] ,\, \card{A}=r}
\plusminus \det Z(M)[I,A] \cdot \det Z(M)[\overline{I},\overline{A}]
\end{equation}
Clearly $A$ must be a subset of the columns of $Q_1$ and also $A \supseteq \partial_1 M$,
otherwise either $Z(M)[I,A]$ or $Z(M)[\overline{I},\overline{A}]$ will contain a zero column.
So \Equation{expansion1} may be rewritten
$$ \det Z(M) ~=~ \sumstack{A \subseteq \overline{\partial_1 M} \\ \card{A}=r-\card{\partial_1 M}}
\plusminus \det Q_1[*,A \union \partial_1 M] \cdot
\det Z(M)[\overline{I},\overline{A \union \partial_1 M}]. $$
We now apply the Laplace expansion to $Z(M)[\overline{I},\overline{A \union \partial_1 M}]$ with the
set of columns $I' = \set{1, \ldots, r}$.
A similar argument yields
\begin{equation}
\EquationName{detexpansion}
\det Z(M) ~= \!\!\!
\sumstack{A \subseteq \overline{\partial_1 M} \\ \card{A}=r-\card{\partial_1 M}} \!\!\!
\plusminus \det Q_1[*,A \union \partial_1 M] ~\cdot\!\!\!\!
\sumstack{B \subseteq \overline{\partial_2 M} \\ \card{B}=r-\card{\partial_2 M}} \!\!\!
\plusminus \det Q_2[B \union \partial_2 M,*]
\cdot \det X'[\overline{B},\overline{A}]
\end{equation}

Consider now the matrix $X'[\overline{B},\overline{A}]$.
We may use the remaining entries of $D_1$
to clear their rows, and similarly for $D_2$. The resulting matrix has the form
\begin{align*}
X'' = \begin{pmatrix}
     & D_2' \\
D_1' & \\
     &      & X[A,B] & X[A,C_2 \union S'] \\
     &      & X[C_1 \union S',B] & X[C_1 \union S',C_2 \union S']
\end{pmatrix}
\end{align*}
where $D_1' = \minor{(D_1)}{A}{A}$ and $D_2' = \minor{(D_2)}{B}{B}$.
To simplify our notation, let $Y(A,B)$ denote $X[A \union C_1 \union S' ,\: B \union C_2 \union S']$.
Clearly $\det X'[\overline{B},\overline{A}] = \plusminus \det D_1' \cdot \det D_2' \cdot \det Y(A,B)$.

The factor $\det D_1' \cdot \det D_2'$ is a product of distinct indeterminates.
The crucial observation is that this these products are distinct for each choice of $A$ and $B$.
Therefore there can be no cancellation among the terms of \Equation{detexpansion},
so it follows that $Z(M)$ is non-singular iff there exist
$A \subseteq \overline{\partial_1 M}$ and $B \subseteq \overline{\partial_2 M}$ such that
$Q_1[*,A \union \partial_1 M]$, $Q_2[B \union \partial_2 M,*]$, and $Y(A,B)$ are all non-singular.
Equivalently, we must have $A \union C_1 \in \cB_{\mat_1'}$, $B \union C_2 \in \cB_{\mat_2'}$,
and $Y(A,B)$ non-singular.
By \Lemma{geelen}, this is precisely the condition that $G(M)$ has a bpm.
\end{proof}

By the discussion of \Section{contracted},
an equivalent statement of \Theorem{matchingrank} is as follows.

\begin{corollary}
$M$ is an extensible set iff $Z(M)$ is non-singular.
\end{corollary}

%%%%%%%%%%%%%%%%%%%%%%%%%%%%%%%%%%%%%%%%%%%%%%%%%%%%%%%%%%%%%%%%%%%%%%%%%%%%%%%%%%%%%%%%%%%%%%%%%%%%%
\subsection{Allowed Edges}
\SectionName{allowed}

Suppose that $M$ is an extensible set.
We say that an edge $e=\set{i,j}$ in $G(M)$ is \newterm{allowed} (relative to $M$)
if $M+e$ is also extensible.
This section explains how to test efficiently whether $e$ is allowed.
It will be convenient to identify the vertices $T_1' \union S'$ (vertices $T_2' \union S'$) with
rows (columns) of the submatrix of $X$ in $Z(M)$.
Also, let $N = Z(M)^{-1}$. There are two cases.

\textbf{(A):} Either $i$ or $j$ is in $T_1' \union T_2'$,
say $i \in T_1'$ and $j \in T_2' \union S'$.
Then $Z(M+e) = \minor{Z(M)}{i}{j}$,
modulo a permutation of the rows and columns.
Therefore, $e$ is allowed iff $N_{j,i} \neq 0$,
since $\det Z(M+e) = \plusminus \det \minor{Z(M)}{i}{j}
= \plusminus \det Z(M) \cdot N_{j,i}$.
Furthermore, the matrix $Z(M+e)^{-1}$ may easily be computed from $N$.
It is known (e.g., \cite[Theorem 3.1]{Mucha}) that
\begin{equation}
\EquationName{rank1update}
Z(M+e)^{-1}
~=~
\minor{
\Big(
N \:-\:
N_{*,i}
\cdot
(N_{j,i})^{-1}
\cdot
N_{j,*}
\Big)
}{j}{i}.
\end{equation}
This is a rank-1 update and in fact \Equation{rank1update}
can easily be derived from \Equation{rank1}.

\textbf{(B):} Both $i$ and $j$ are in $S'$.
Then $Z(M+e) = \minor{Z(M)}{\set{i,j}}{\set{i,j}}$.
Jacobi's theorem
%\cite[Section 42]{Aitken}
(see, e.g., Gantmacher \cite[\S 1.4]{Gantmacher})
implies that
$\det \minor{Z(M)}{\set{i,j}}{\set{i,j}}
= \plusminus \det \: N[\set{i,j},\set{i,j}]$.
Therefore, $e$ is allowed iff $\det N[\set{i,j},\set{i,j}] \neq 0$.
As before, the matrix $Z(M+e)^{-1}$ may easily be computed from $N$.
A simple extension of \Equation{rank1update} shows that
\begin{equation}
\EquationName{rank2update}
Z(M+e)^{-1} ~=~
\minor{
\Big(
N \:-\:
N_{*,\set{i,j}}
\cdot
(N_{\set{i,j},\set{i,j}})^{-1}
\cdot
N_{\set{i,j},*}
\Big)
}{\set{i,j}}{\set{i,j}}.
\end{equation}

%%%%%%%%%%%%%%%%%%%%%%%%%%%%%%%%%%%%%%%%%%%%%%%%%%%%%%%%%%%%%%%%%%%%%%%%%%%%%%%%%%%%%%%%%%%%%%%%%%%%%
%%%%%%%%%%%%%%%%%%%%%%%%%%%%%%%%%%%%%%%%%%%%%%%%%%%%%%%%%%%%%%%%%%%%%%%%%%%%%%%%%%%%%%%%%%%%%%%%%%%%%
%%%%%%%%%%%%%%%%%%%%%%%%%%%%%%%%%%%%%%%%%%%%%%%%%%%%%%%%%%%%%%%%%%%%%%%%%%%%%%%%%%%%%%%%%%%%%%%%%%%%
\section{Basic Path-Matching Algorithm}
\SectionName{bpmalg}

In this section we develop an efficient algorithm for the basic path-matching problem.
Let $G=(T_1,T_2,S,E,\mat_1,\mat_2)$ be an instance, and let $n = \card{T_1} + \card{T_2} + \card{S}$.
We use the following general approach.
We start with the matrix $Z$ and randomly substitute values for the indeterminates
from the field $\bF$, or a sufficiently large extension.
Operating in an extension field increases the running time by a factor of
$O( \log n \cdot \text{poly}( \log \log n ) )$.
By standard arguments, which we omit, this random substitution does not affect
the rank of $Z$ with high probability.
If $Z$ is singular then \Theorem{matchingrank} implies that no basic path-matching exists (whp).
So assume that $Z$ is non-singular.
The algorithm computes $Z^{-1}$, which requires only $O(n^\omega)$ time.
Let $N$ be the transpose of the
lower-right submatrix of $Z^{-1}$ (corresponding to the matrix $X$).

The algorithm searches for allowed edges, following the approach of \Section{allowed}.
Once an allowed edge $\set{i,j}$ is found, we set $M=\set{\, \set{i,j} \,}$ and
construct a bpm for the contracted instance $G(M)$.
This requires constructing the corresponding matrix $Z(M)^{-1}$,
which can be done via the updates in \Equation{rank1update} and \Equation{rank2update}.
Performing these updates immediately takes $O(n^2)$ time and would lead to an $O(n^3)$ algorithm.
Instead, we will store the parameters of the updates that are to be performed
and only apply the updates to portions of $N$ that will be examined soon.
An update that has yet to be applied is called \emph{pending}.
A submatrix for which no updates are pending is called \emph{clean}.

Each update involves subtracting from $Z(M)^{-1}$ the product of three matrices $u$, $c$ and $v$
where $u$ is $n \times k$, $c$ is $k \times k$, $v$ is $k \times n$,
and $k \in \set{1,2}$.
Furthermore, $u$, $c$ and $v$ are submatrices of $Z(M)^{-1}$ (at the time of the update),
and the submatrices involved in different updates are disjoint.
Therefore we may represent all updates by three $n \times n$ matrices
$U$, $C$ and $V$, and two lists $\pi_c$ and $\pi_r$.
$U$ stores the $u$-matrices,
$C$ stores the $c$-matrices, and $V$ stores the $v$-matrices;
they are all initially zero.
The list $\pi_c$ specifies which columns of $U$ and $C$
are involved in each update, and $\pi_r$ similarly describes the rows of $V$ and $C$.

The algorithm proceeds recursively,
starting from the graph $G$ with vertex sets $T_1$, $T_2$ and $S$.
For convenience, we imagine augmenting the submatrix $X$ so that its
row-indices and column-indices are identical:
we add empty columns corresponding to $T_1$ and empty rows corresponding to $T_2$.
Next, we arbitrarily partition $T_1 \union T_2 \union S$ into $\alpha \geq 3$ equal-sized
parts $V_1, \ldots, V_\alpha$, where the value of $\alpha$ will be chosen
later. For each unordered pair of parts $\set{V_a,V_b}$,
we recurse on the subproblem with vertex set $V_a \union V_b$.
A leaf of the recursion considers a $2 \times 2$ submatrix $N[\set{i,j},\set{i,j}]$;
furthermore, every such submatrix is considered by at least one leaf.

The algorithm maintains the invariant that the submatrix
$N[V_a \union V_b,V_a \union V_b]$ is clean whenever it starts or
finishes solving the subproblem for vertex set $V_a \union V_b$.
Consider the a leaf of the recursion tree for a pair of vertices $\set{i,j}$.
As in \Section{allowed} there are two cases.
\textbf{(A):} $i \in T_1$. The edge $\set{i,j}$ is allowed iff $N_{i,j} \neq 0$.
The case where $j \in T_2$ is analogous.
\textbf{(B):}
Both $i$ and $j$ belong to $S$.
The edge $\set{i,j}$ is allowed iff $\det N[\set{i,j},\set{i,j}] \neq 0$.

If $\set{i,j}$ is allowed, we perform the following steps.
First, we adjoin $\set{i,j}$ to $M$.
Next, the parameters of the update (either \Equation{rank1update} or \Equation{rank2update},
depending on whether case \textbf{(A)} or \textbf{(B)} applies)
are stored in the matrices $U$, $C$ and $V$;
however, matrices $U$ and $V$ are updated only conceptually.
To satisfy our invariant we need only ensure that the submatrices of $U$, $V$ and
$N$ corresponding to the \emph{current} subproblem are updated.
The lists $\pi_c$ and $\pi_r$ are also updated to indicate which
columns and rows are involved in this update.

Suppose that the subproblem with vertices $V_a \union V_b$ has just been processed,
and now we continue processing its ``parent'' in the recursion tree, which we denote by $G$.
Any update generated during processing of the subproblem is pending.
Suppose that $N$ has the block form
$N = \begin{smallpmatrix}
N_{NW} & N_{NE} \\
N_{SW} & N_{SE}
\end{smallpmatrix}$ where, without loss of generality, the vertices
$V_a \union V_b$ correspond to the ``north-west'' submatrix.
Assume that $U$, $C$ and $V$ have the same block form.
The submatrix $N_{NW}$ is clean by our invariant but the other three submatrices of $N$ are not,
nor are $U_{SW}$ and $V_{NE}$.
To update $U_{SW}$, let $\tU=N_{SW}$, $\tC=C_{NW}$, and let $\tV$ be a copy of $V_{NW}$ where rows
that do \emph{not} correspond to pending updates are set to zero.
Assume that these matrices are ordered according to $\pi_c$ and $\pi_r$.
Observe that $\tV$ is strictly upper triangular,
since columns participating in an update are set to zero.

To compute $U_{SW}$, we will employ two tricks.
The matrix $\tC$ is a block-diagonal matrix with $1 \times 1$ and $2 \times 2$ blocks.
The $2 \times 2$ blocks correspond to updates of type \textbf{(B)}, which are rank-2 updates.
Such an update can be unfurled into four rank-1 updates as follows:
\begin{equation}
\EquationName{unfurl}
\setlength{\arraycolsep}{1.5pt}
\renewcommand{\arraystretch}{0.88}
\begin{pmatrix}
\vline & \vline \\
 u_1 & u_2 \\
\vline & \vline
\end{pmatrix}
\cdot
\begin{pmatrix}
c_{1,1} & c_{1,2} \\
c_{2,1} & c_{2,2}
\end{pmatrix}
\cdot
\begin{pmatrix}
\myhline & v_1 & \myhline \\
\myhline & v_2 & \myhline
\end{pmatrix}
=
\begin{pmatrix}
\vline & \vline & \vline & \vline \\
 u_1 & u_2 & u_2 & u_1 \\
\vline & \vline & \vline & \vline
\end{pmatrix}
\cdot
\begin{pmatrix}
c_{1,1} \\
 & c_{2,2} \\
 & & c_{1,2} \\
 & & & c_{2,1}
\end{pmatrix}
\cdot
\begin{pmatrix}
\myhline & v_1 & \myhline \\
\myhline & v_2 & \myhline \\
\myhline & v_1 & \myhline \\
\myhline & v_2 & \myhline
\end{pmatrix}
\end{equation}
We now wish to apply all of these updates.
The difficulty is that the columns of $\tU$ involved in the $j\th$ update may be modified
by the $i\th$ update if $i < j$.
To surmount this difficulty, we use the following lemma.

\begin{lemma}[Sequential Update Lemma]
\LemmaName{sequpdate}
Let $X$ and $Y$ be $n \times n$ matrices where $Y$ is strictly upper triangular.
Define a sequence of matrices by $X^{(0)} = X$ and
$X^{(i)} = X^{(i-1)} + X^{(i-1)}_{*,i} \cdot Y_{i,*}$ for $1 \leq i \leq n$.
Let $X \otimes Y$ denote the matrix $X^{(n)}$.
Then $X \otimes Y = X \cdot (I-Y)^{-1}$.
\end{lemma}
\begin{proof}
First, note that $X^{(j)}_{*,i} = X^{(k)}_{*,i}$ if $i \leq j \leq k$
since $Y$ is strictly upper triangular.
Define
$$ A=\begin{pmatrix} I-Y & 0 \\ X & I \end{pmatrix}, $$
and consider performing Gaussian elimination on $A$.
Let $S^{(i)}$ denote the south-west submatrix of $A$ just before the $i\th$ elimination.
An easy inductive argument shows that $S^{(i)}_{*,\,i:n} = X^{(i)}_{*,\,i:n}$.
The (lower half of the) column vector involved in the $i\th$ elimination is therefore
$S^{(i)}_{*,\,i}=X^{(n)}_{*,\,i}$.
Now consider the LU-decomposition of $A$:
$$
\begin{pmatrix} I-Y & 0 \\ X & I \end{pmatrix}
=
\begin{pmatrix} I & 0 \\ B & I \end{pmatrix}
\cdot
\begin{pmatrix} I-Y & 0 \\ 0 & I \end{pmatrix}.
$$
It is well-known that $B_{*,i}$ is precisely the (lower half of the) column involved
in the $i\th$ elimination (see, e.g., Strang \cite{Strang}).
Thus $B = X^{(n)} = X \otimes Y$.
The lemma follows by observing that $X = B \cdot (I-Y)$.
\end{proof}

We can therefore compute $N_{SW} = \tU \otimes (-\tC \cdot \tV)$ in $O(n^\omega)$ time
by the sequential update lemma.
The columns from $N_{SW}$ corresponding to pending updates are then copied into
$U_{SW}$ and set to zero.
This makes $N_{SW}$ and $U_{SW}$ clean.
A symmetric argument explains how to make $V_{NE}$ and $N_{NE}$ clean.
Finally, let $\tU$, $\tC$ and $\tV$ denote the submatrices of
$U_{SW}$, $C_{NW}$ and $V_{NE}$ corresponding to the pending updates.
We make $N_{SE}$ clean by setting $N_{SE} = N_{SE} - \tU \tC \tV$
(cf.~\Equation{rank1update} and \Equation{rank2update}).
All matrices (within the current subproblem) are now clean,
and the recursion may continue.

Starting from an instance with $n$ vertices,
this algorithm recurses on $\binom{\alpha}{2}$ subproblems with $\frac{2n}{\alpha}$ vertices.
After solving each subproblem, the algorithm ensures that the matrices
of the parent subproblem are clean.
So the time required satisfies the recurrence
\begin{equation}
\EquationName{recurrence}
T(n) = \smallbinom{\alpha}{2} \cdot T\Big(\smallfrac{n}{\alpha/2}\Big)
+ O\Big( \smallbinom{\alpha}{2} \cdot n^\omega \Big).
\end{equation}
The solution of the recurrence is $T(n)=O(n^\omega)$ if $\alpha$ is a constant
chosen such that $\log_{\alpha/2} \binom{\alpha}{2} < \omega$.
Since $\log_{\alpha/2} \binom{\alpha}{2} < 2 + \frac{1}{\log \alpha-1}$,
there exists an appropriate choice of $\alpha$, assuming that $\omega>2$.
If $\omega=2$, this analysis gives an $O(n^{2+\epsilon})$ bound, for any $\epsilon > 0$.
We may obtain an improved bound by modifying the algorithm to keep a bit vector to
ensure that each recursive subproblem is entered at most once.
Thus at level $i$ of the recursion there will be
$\binom{ \alpha (\alpha/2)^i }{ 2 } < 2 (\alpha/2)^{2i+2}$ subproblems,
and the time for applying updates is $O(( n (\alpha/2)^{-i+1} )^\omega)$.
Since there are at most $\log_{\alpha/2} n$ levels of recursion,
this yields a bound of $O(n^\omega \log n)$ which is valid even if $\omega=2$.

%%%%%%%%%%%%%%%%%%%%%%%%%%%%%%%%%%%%%%%%%%%%%%%%%%%%%%%%%%%%%%%%%%%%%%%%%%%%%%%%%%%%%%%%%%%%%%%%%%%%%
%%%%%%%%%%%%%%%%%%%%%%%%%%%%%%%%%%%%%%%%%%%%%%%%%%%%%%%%%%%%%%%%%%%%%%%%%%%%%%%%%%%%%%%%%%%%%%%%%%%%%
%%%%%%%%%%%%%%%%%%%%%%%%%%%%%%%%%%%%%%%%%%%%%%%%%%%%%%%%%%%%%%%%%%%%%%%%%%%%%%%%%%%%%%%%%%%%%%%%%%%%%
\section{Applications}

In this section we discuss several applications of basic path-matchings and
our preceding results.

\subsection{Non-bipartite Matching}

The problem of finding a perfect matching in a general graph $G$ is the special case
of the basic path-matching problem where $T_1=T_2=\emptyset$.
Let $n = \card{S}$ be the number of vertices in the graph.
In this case, an extensible set $M$ is a matching that is contained in a perfect matching.
The matrix $Z(M)$ is simply the Tutte matrix restricted to the vertices exposed by $M$,
and is therefore skew-symmetric.
This fact has useful implications.

First of all, we may wish to find a maximum matching instead of a perfect matching.
To do so, we first choose a random substitution for the indeterminates in $Z$,
then find a full-rank submatrix, say $Z[A,B]$, by the Hopcroft-Bunch \cite{Bunch} algorithm.
Lov\'asz \cite{RandomMatching} showed that $\card{A}$ is twice the cardinality of a maximum matching.
Furthermore, it is known that $Z[A,A]$ also has full rank
(by Frobenius' theorem \cite{RabinVazirani} or the Grassman-Pl\"{u}cker identity \cite[p434]{Murota}).
This implies that $G[A]$ has a perfect matching,
so we may apply the algorithm of \Section{bpmalg} to $G[A]$.

Since $Z(M)$ is skew-symmetric, the matrix $N=Z(M)^{-1}$ is as well \cite[p521]{Cullis}.
Therefore, every $2 \times 2$ submatrix inspected by
our algorithm is of the form $\begin{smallpmatrix} 0 & x \\ -x & 0 \end{smallpmatrix}$.
Consequently, \Equation{rank2update} simplifies to the sum of two rank-1 updates,
and the transformation of \Equation{unfurl} becomes unnecessary.
Moreover, this submatrix has non-zero determinant iff $x \neq 0$.
This simplifies the test of \Section{allowed} \textbf{(B)} and
also yields an alternative proof of \cite[Lemma 3]{RabinVazirani}.

%%%%%%%%%%%%%%%%%%%%%%%%%%%%%%%%%%%%%%%%%%%%%%%%%%%%%%%%%%%%%%%%%%%%%%%%%%%%%%%%%%%%%%%%%%%%%%%%%%%%
\subsection{Bipartite Matroid Matchings}

An instance of the bipartite matroid matching problem is
a tuple $G=(T_1,T_2,E,\mat_1,\mat_2)$ where $E \subseteq T_1 \times T_2$
and $\mat_i=(T_i,\cI_i,r_i)$ is a matroid.
An \newterm{independent matching} in $G$ is a matching $M \subseteq E$ such that
$\partial_1 M \in \cI_1$ and $\partial_2 M \in \cI_2$.
The objective is to find a maximum cardinality independent matching.
The bipartite matroid matching problem is clearly a special case of the basic perfect-matching
problem where $S=\emptyset$.
\Theorem{matchingrank} extends to show that
the maximum cardinality of an independent matching in $G(M)$ is $\rank Z(M) - 2n$.
(Murota's characterization of the rank of mixed matrices \cite{Murota} is convenient for proving
this extension.)

Additionally, the algorithm of \Section{bpmalg} immediately solves the bipartite matroid matching
problem in $\tO(n^\omega)$ time.
When applied to bipartite matroid matching problems, this algorithm
amounts to executing the
Mucha-Sankowski bipartite matching algorithm on the lower-right submatrix of $Z^{-1}$.
Thus we have the following surprising result:
when the Mucha-Sankowski algorithm executes on $X$ and $X^{-1}$,
it computes a maximum bipartite matching.
However, when it executes on $X$ and $\tQ = Q_2 (Q_1 X Q_2)^{-1} Q_1$
(which is the lower-right submatrix of $Z^{-1}$)
it computes a maximum independent matching.

A somewhat more efficient bound may be obtained by considering the rank $r$.
The algorithm begins by considering the initial graph $G$ and $\tQ$.
We note that $\tQ$ may be computed in $O(n^2 r^{\omega-2})$ time.
The vertex sets $V_1$ and $V_2$ are each partitioned into two parts.
The algorithm recurses on all four ordered pairs of parts,
and the corresponding submatrices of $\tQ$.
Since since any independent matching has size at most $r$,
the number of subproblems per level that generate an update is at most $r$.
Furthermore, the size of the updates computed may also be parameterized by $r$.
These observations lead to the bound $\tO(n^2 r^{\omega-2} + r^{\omega + 1})$
which is tighter than $\tO(n^\omega)$ when $r = o(n^{\omega/(\omega+1)})$.
The details are mundane and hence omitted.

%%%%%%%%%%%%%%%%%%%%%%%%%%%%%%%%%%%%%%%%%%%%%%%%%%%%%%%%%%%%%%%%%%%%%%%%%%%%%%%%%%%%%%%%%%%%%%%%%%%%
\subsection{Matroid Intersection}

A matroid intersection problem is a bipartite matroid matching problem
where the edges $E$ in the bipartite graph are a matching.
Alternatively, if $\mat_1 = (V,\cI_1)$ and $\mat_2 = (V,\cI_2)$ are two matroids,
a set $J \subseteq V$ is called an \newterm{intersection}
if $J \in \cI_1 \intersect \cI_2$.
A maximum intersection is one with maximum cardinality.
If $J \subseteq V$ is contained in a maximum cardinality intersection, it is called extensible.
If $J$ is extensible, $i \not \in J$, and $J+i$ is also extensible
then element $i$ is called allowed.

For matroid intersection problems, the matrix $Z(M)$ can be greatly simplified,
owing to the fact that all principal submatrices of $X$ are non-singular.
If $J$ is an extensible intersection, we define the matrix $Z(J)$ as
$$ Z(J) = \begin{pmatrix}
              & Q_1^{J} & Q_1^{\bar{J}} \\
Q_2^{J}       &         &  \\
Q_2^{\bar{J}} &         & \minor{X}{J}{J}
\end{pmatrix}.
$$
By an argument similar to \Theorem{matchingrank}, it follows that
the cardinality of a maximum intersection in $\mat_1 / J$ and $\mat_2 / J$
is $\rank Z(J) + \card{J} - \card{V}$.
This result was mentioned by Geelen \cite{GeelenCourse}
(and attributed to Murota \cite{Murota}) for the special case that $J = \emptyset$.

The characterization of allowed elements for matroid intersection problems is as follows.
We identify the elements of $V$ with the
rows and columns of the submatrix of $X$ in $Z(M)$.

\begin{lemma}
\LemmaName{invdiag}
Let $J$ be an extensible intersection and let $i \in V \setminus J$.
The element $i$ is allowed if and only if $(Z(J)^{-1})_{i,i} \neq z_i^{-1}$.
\end{lemma}
\begin{proof}
As mentioned above, $J+i$ is extensible if and only if $Z(J+i)$ is non-singular.
By linearity of the determinant, $\det Z(J+i) = \det Z(J) - z_i \cdot \det \minor{Z(J)}{i}{i}$.
Since $(Z(J)^{-1})_{i,i} = \det \minor{Z(J)}{i}{i} / \det Z(J)$,
we have $\det Z(J+i) = \det Z(J) \cdot (1 - z_i \cdot (Z(J)^{-1})_{i,i} )$.
Thus $\det Z(J+i) = 0 \iff Z(J)^{-1}_{i,i} = z_i^{-1}$.
\end{proof}

The structure of $Z(J)$ will play a helpful role in our algorithms for matroid intersection below.
The Schur complement of $\minor{X}{J}{J}$ is
$$
Y(J) \defeq
\begin{pmatrix}
-Q_1^{\bar{J}} \cdot X(J)^{-1} \cdot Q_2^{\bar{J}}   &    ~~Q_1^J  \\
Q_2^J
\end{pmatrix}.
$$
For simplicity, let $Y$ denote $Y(\emptyset)$.
One may verify that
\begin{align}
\EquationName{zjinv}
Z(J)^{-1} =
\begin{pmatrix}
Y(J)^{-1} &
- (Y(J)^{-1})_{*,\,[r]} \cdot Q_1^{\bar{J}} \cdot X(J)^{-1} \\
- X(J)^{-1} \cdot Q_2^{\bar{J}} \cdot (Y(J)^{-1})_{[r],\,*} &
~X(J)^{-1} +
X(J)^{-1} \cdot Q_2^{\bar{J}} \cdot (Y(J)^{-1})_{[r],\,[r]} \cdot Q_1^{\bar{J}} \cdot X(J)^{-1}
\end{pmatrix}.
\end{align}
The matroid intersection algorithms developed below will use this structure
of $Z(J)^{-1}$ crucially.

%%%%%%%%%%%%%%%%%%%%%%%%%%%%%%%%%%%%%%%%%%%%%%%%%%%%%%%%%%%%%%%%%%%%%%%%%%%%%%%%%%%%%%%%%%%%%%%%%%%%
\subsubsection{Matroid intersection algorithm I}

In this section we describe our first matroid intersection algorithm, which achieves
running time $O(n r^{\omega-1} \log r)$.
We start with the matrix $Z$ and randomly substitute values for the indeterminates
from the field $\bF$, or a sufficiently large extension.
Next, we restrict our attention to a full-rank principal submatrix of the form $Z[A,A]$
with $\set{r+1,\ldots,r+n} \subseteq A$.
Such a submatrix may be found in $O(mn^{\omega-1})$ time by the following scheme.
First compute $Y$, which requires $O(nr^{\omega-1})$ time.
Now, since $Y$ is the Schur complement of $X$,
it follows that $\rank Z = \rank Y + n$.
Thus the submatrices of $Y$ of rank $k$
correspond to submatrices of $Z$ with the desired form and rank $k+n$.
A maximum rank submatrix of $Y$ can be found in $O(r^\omega)$ time
by performing Gaussian elimination.

The present algorithm maintains an extensible matching, initially empty,
and searches for allowed edges using \Lemma{invdiag}.
This seemingly requires computing the entire matrix $Z^{-1}$,
which would require a prohibitive $\Omega(n^2)$ time.
Initially we work under the assumption that $Z^{-1}$ has been completely computed,
and later we will show that this assumption can be removed.
We also assume that both $n$ and $r$ are powers of two.

Suppose that an allowed element $i$ has been found and we now wish to construct $Z(\set{i})^{-1}$.
The matrix $Z(\set{i})$ is identical to $Z$ except that $z_i$ has been set to $0$.
This can be expressed as the rank-1 update $Z(\set{i}) = Z + (-z_i) e_{r+i} e_{r+i}\transpose$.
The corresponding change to the inverse matrix is
\begin{equation}
\EquationName{matroidupdate}
Z(\set{i})^{-1} = Z^{-1} - (z_i^{-1}+(Z^{-1})_{r+i,r+i})^{-1} \: (Z^{-1})_{*,r+i} \: (Z^{-1})_{r+i,*}.
\end{equation}
Because the updates have this form, we can again adopt the approach
of \Section{bpmalg}.
To harmonize the notation, let $N$ denote the lower-right submatrix $(Z^{-1})_{r+1:r+n,\,r+1:r+n}$.
As before, updates will be represented using matrices $U$, $C$ and $V$.

The algorithm is recursive, initially considering the entire vertex set $V$
and the entire submatrix $N$.
The vertex set is partitioned: let $V_1$ be the first $n/2$ vertices and $V_2$ the remainder.
The algorithm recurses first on the subproblem $V_1$, and then on $V_2$.
However, the invariant used in \Section{bpmalg} does not lead to an efficient algorithm
for matroid intersection.
We obtain an efficient algorithm through the following
more relaxed invariants:
\begin{enumerate}
\item When the recursion enters a subproblem with vertex set $A$
then $N[A,A]$ is clean if $\card{A} \leq r$.
If $\card{A} > r$ then the $r \times r$ blocks on the diagonal of $N[A,A]$ are clean.
\item For any subproblem, before its second child subproblem begins processing,
then the $U_{SW}$ and $V_{NE}$ submatrices are clean.
\end{enumerate}
A leaf of the recursion tree considers only a single vertex,
and simply checks if $N_{i,i} \neq z_i^{-1}$.
If so, the element $i$ is added to the intersection,
and we conceptually perform the update of \Equation{matroidupdate}.
To represent the update, we store its scalar parameter in $C_{i,i}$.

Consider a subproblem at level $i$ of the recursion tree.
The number of vertices in this subproblem is $n 2^{-i}$.
Once its first child has finished, we must update $U_{SW}$ and $V_{NE}$.
This step is more involved than in \Section{bpmalg} because of our
relaxed invariants.
We say that an update is \newterm{old} if it was produced \emph{before} entering subproblem $V_1$.
The difficulty is that old updates might not have been applied to $N_{SW}$.
The easy case is when $n 2^{-i} \leq r$.
In this case, invariant (1) implies that the old updates have been applied.
It remains to apply the pending updates that were generated within the first
child subproblem.
This is done as in \Section{bpmalg}, using only $O( (n 2^{-i})^\omega )$ time.

The more difficult case is when $n 2^{-i} > r$.
In this case, there may be old updates which have not yet been applied to $N_{SW}$.
In fact, no updates have been applied to $N_{SW}$ whatsoever.
However, the portions of $U$ and $V$ which contain the updates relevant to $N_{SW}$ are clean;
this follows from applying invariant (2) to the parent of the current subproblem.
The only portion of $N_{SW}$ that is of interest is
the submatrix corresponding to columns where updates were made.
Let us denote this submatrix by $\tN$.
Its size is at most $n 2^{-i} \times r$.
The number of old updates which must be applied to $\tN$ is also at most $r$.
Therefore the time required to apply the old updates is bounded by the time to multiply
an $n 2^{-i} \times r$ matrix by an $r \times r$ matrix,
which is $O(n 2^{-i} r^{\omega - 1})$ time.
The updates generated within the first child subproblem can also be applied to $N_{SW}$
in $O(n 2^{-i} r^{\omega - 1})$ time by the sequential update lemma,
as in \Section{bpmalg}.
A similar argument applies to updating $N_{NE}$ and $V_{NE}$.
Finally, we must update $N_{SE}$.
To restore the invariants, we only need update its $r \times r$ diagonal blocks.
This requires time only $O( n 2^{-i} r^{\omega-1} )$ by the obvious approach.

To analyze the time required by this algorithm, recall that
there are $2^i$ subproblems at level $i$.
For levels $i < \log(n/r)$, the total time required is
$ \sum_{i=1}^{\log (n/r)} 2^{i} \cdot O( n 2^{-i} r^{\omega-1} ) = O( n r^{\omega-1} \log n )$.
For levels $i \geq \log(n/r)$, the time is
\begin{align*}
&\sum_{i=\log (n/r)}^{\log n} 2^{i} \cdot O( (n 2^{-i})^\omega )
= \sum_{i=\log (n/r)}^{\log n} O( n^\omega 2^{-(\omega-1)i} )
= \sum_{i=0}^{\log r} O( n 2^{(\omega-1)i} )
= O( n r^{\omega-1} ),
\end{align*}
so the total time required is $\tO(n r^{\omega-1})$.

The preceding discussion assumes that $N$ was initially fully computed.
However, it is clear that the only parts of $N$ that are needed are those
in the $r \times r$ diagonal blocks and those that are involved in updates.
It is straightforward to extend the algorithm so that the necessary parts of
$N$ are computed on demand.
At the beginning of the algorithm, we compute only the $r \times r$ diagonal blocks of $N$.
As shown by \Equation{zjinv}, this requires multiplying $Q_2 \transpose Y^{-1} Q_1$
plus some negligible additional work, and hence $O(n r^{\omega-1} )$ time is required.
Next, consider performing an update where $n 2^{-i} > r$.
Then we must compute the initial submatrix $\tN$ (before any updates are applied).
Recall that $\tN$ has size at most $n 2^{-i} \times r$.
Then, refering again to \Equation{zjinv}, we see that $\tN$
can be computed in $O(n 2^{-i} r^{\omega - 1})$ time.
Thus the preceding analysis applies without change.

%%%%%%%%%%%%%%%%%%%%%%%%%%%%%%%%%%%%%%%%%%%%%%%%%%%%%%%%%%%%%%%%%%%%%%%%%%%%%%%%%%%%%%%%%%%%%%%%%%%%
\subsubsection{Matroid intersection algorithm II}
\SectionName{mlarge}

We conclude the paper with a simpler
matroid intersection algorithm that is efficient
if $n$ is sufficiently large, or if we restrict ourselves
to naive matrix multiplication.
As usual, the algorithm chooses random values for the indeterminates in $Z$
from a sufficiently large field extension, and restricts attention to a full-rank submatrix.
The algorithm relies on the fact that the diagonal elements of $Z^{-1}$ can be computed quickly,
and can also be updated as elements are added to the extensible intersection.
For convenience, let $\cY(J)$ denote $(Y(J)^{-1})_{[r],\,[r]}$.
As is evident from \Equation{zjinv}, computing the diagonal elements of $Z(J)^{-1}$
involves computing the matrix $\cY$.
The following lemma shows how adding an element to the intersection changes $\cY$.
It is obvious that the change is a rank-1 update,
but less obvious that the parameters of the update
exclusively involve $\cY$, $Q_1$ and $Q_2$.

\begin{lemma}
\LemmaName{yupdate}
Let $J$ be an intersection and let $i \not \in J$ be such that $J+i$ is extensible.
Let $u = (Q_1)_{*,i}$ and let $v = (Q_2 \transpose)_{*,i}$.
Then
$ \cY(J+i) \:=\: \cY(J) - (v\transpose \, \cY(J) \, u)^{-1}
\cdot (\cY(J) \, u) \cdot (v\transpose \, \cY(J)). $
\end{lemma}
\begin{proof}
Note that $Z(J+i) = Z(J) + (-z_i) e_{r+i} e_{r+i}\transpose$.
Hence
\begin{equation}
\EquationName{zupdate}
Z(J+i)^{-1} ~=~ Z(J)^{-1} -
\alpha^{-1} \cdot
(Z(J)^{-1} \, e_{r+1}) \cdot (e_{r+1}\transpose \, Z(J)^{-1} ),
\end{equation}
where $\alpha = (-z_i^{-1} + (Z(J)^{-1})_{r+i,r+i})$.
Using the structure of $Z(J)^{-1}$ in \Equation{zjinv},
we see that $(Z(J)^{-1})_{r+i,r+i} = z_i^{-1} ( 1 + z_i^{-1} v\transpose \cY u z_i^{-1})$,
so $\alpha = z_i^{-2} v \transpose \cY u$.
Now, restricting \Equation{zupdate} to the first $r$ rows and columns, we obtain
$$ \cY(J+1) ~=~ \cY(J) - \alpha \cdot
(Z(J)^{-1})_{1:r,r+i} \cdot (Z(J)^{-1})_{r+i,1:r}.
$$
Again using the structure of $Z(J)^{-1}$,
we see that $(Z(J)^{-1})_{1:r,r+i}$ is the $(i-j)\th$ column of
$-(Y^{-1})_{1:r,1:r} \cdot Q_1^{\bar{J}} \cdot \tZ(J)^{-1}$,
which is $-z_i^{-1} \cY \, u$.
Similarly, $(Z(J)^{-1})_{r+i,1:r} = -z_i^{-1} v \cY$,
establishing the lemma.
\end{proof}

The algorithm begin with $J=\emptyset$ and computes the initial matrix $\cY(J)=Y^{-1}$,
which requires $O(n r^{\omega-1})$ time, as observed above.
Using $\cY(J)$, we may efficiently compute the diagonal elements of $Z(J)^{-1}$ in
$r \times r$ blocks.
Each block requires $O(r^\omega)$ time.
Computing the first diagonal block allows us to determine if an element
$i$ among the first $r$ is allowed, using \Lemma{invdiag}.
If so, we compute $\cY(J+i)$ using \Lemma{yupdate}
and recompute the current block so that it becomes a submatrix of
$Z(J+i)^{-1}$ rather than $Z(J)^{-1}$.
As is evident from \Lemma{yupdate} and \Equation{zjinv},
both of these computations are rank-1 updates and hence require only $O(r^2)$ time.
The algorithm adjoins $i$ to $J$ and repeats this process until $\card{J} = r$.

We now analyze the time required by this algorithm.
Computing a block of $r$ diagonal elements of $Z(J)^{-1}$ requires only $O(r^\omega)$ time.
Since there are $\ceil{n/r}$ such blocks, the total time for these computations is
$O(n r^{\omega-1})$.
After adding each element to $J$, we compute $\cY(J+i)$ and recompute the current block
in $O(r^2)$ time.
Since $\card{J}=r$ at the completion of the algorithm, the total time for these updates is $O(r^3)$.
Thus the running time of the algorithm is $O(n r^{\omega-1} + r^3)$.
This bound is
$O(n r^{\omega-1})$ for $n = \Omega(r^{4-\omega})$.
If only naive matrix multiplication is used, and the field size is sufficiently large,
we obtain a bound of $O(n r^2)$.

%%%%%%%%%%%%%%%%%%%%%%%%%%%%%%%%%%%%%%%%%%%%%%%%%%%%%%%%%%%%%%%%%%%%%%%%%%%%%%%%%%%%%%%%%%%%%%%%%%%%
\section*{Acknowledgements}

The author would like to thank
Michel Goemans, Satoru Iwata, and David Karger for discussions on this topic.

%%%%%%%%%%%%%%%%%%%%%%%%%%%%%%%%%%%%%%%%%%%%%%%%%%%%%%%%%%%%%%%%%%%%%%%%%%%%%%%%%%%%%%%%%%%%%%%%%%%%
\renewcommand{\baselinestretch}{0.98}
\begin{small}
\bibliography{Notes}

\begin{thebibliography}{10}

\bibitem{Aigner}
M.~Aigner and T.~A. Dowling.
\newblock Matching theory for combinatorial geometries.
\newblock {\em Transactions of the American Mathematical Society},
  158(1):231--245, July 1971.

\bibitem{Aitken}
A.~C. Aitken.
\newblock {\em Determinants and Matrices}.
\newblock Interscience Publishers, New York, ninth edition, 1956.

\bibitem{Barvinok}
A.~I. Barvinok.
\newblock New algorithms for linear $k$-matroid intersection and matroid
  $k$-parity problems.
\newblock {\em Mathematical Programming}, 69:449--470, 1995.

\bibitem{Bunch}
J.~R. Bunch and J.~E. Hopcroft.
\newblock Triangular factorization and inversion by fast matrix multiplication.
\newblock {\em Mathematics of Computation}, 28(125):231--236, 1974.

\bibitem{Camerini}
P.~M. Camerini, G.~Galbiati, and F.~Maffioli.
\newblock Random pseudo-polynomial algorithms for exact matroid problems.
\newblock {\em Journal of Algorithms}, 13(2):258--273, 1992.

\bibitem{Cook}
W.~J. Cook, W.~H. Cunningham, W.~R. Pulleyblank, and A.~Schrijver.
\newblock {\em Combinatorial Optimization}.
\newblock Wiley, 1997.

\bibitem{Coppersmith}
D.~Coppersmith and S.~Winograd.
\newblock Matrix multiplication via arithmetic progressions.
\newblock {\em Journal of Symbolic Computation}, 9(3):251--280, 1990.

\bibitem{Cullis}
C.~E. Cullis.
\newblock {\em Matrices and Determinoids}, volume~2.
\newblock Cambridge University Press, 1918.

\bibitem{Cunningham}
W.~H. Cunningham.
\newblock Improved bounds for matroid partition and intersection algorithms.
\newblock {\em SIAM Journal on Computing}, 15(4):948--957, Nov. 1986.

\bibitem{CunninghamGeelenManuscript}
W.~H. Cunningham and J.~F. Geelen.
\newblock Vertex-disjoint directed paths and even circuits.
\newblock Manuscript.

\bibitem{CunninghamGeelenFOCS}
W.~H. Cunningham and J.~F. Geelen.
\newblock The optimal path-matching problem.
\newblock In {\em Proceedings of the 37th Annual IEEE Symposium on Foundations
  of Computer Science (FOCS)}, pages 78--85, 1996.

\bibitem{CunninghamGeelen}
W.~H. Cunningham and J.~F. Geelen.
\newblock The optimal path-matching problem.
\newblock {\em Combinatorica}, 17(3):315--337, 1997.

\bibitem{EdmondsSubmodular}
J.~Edmonds.
\newblock Submodular functions, matroids, and certain polyhedra.
\newblock In R.~Guy, H.~Hanani, N.~Sauer, and J.~Sch{\"o}nheim, editors, {\em
  Combinatorial Structures and Their Applications}, pages 69--87. Gordon and
  Breach, 1970.
\newblock Republished in M. J{\"{u}}nger, G. Reinelt, G. Rinaldi, editors,
  \emph{Combinatorial Optimization -- Eureka, You Shrink!}, Lecture Notes in
  Computer Science 2570, pages 11--26. Springer-Verlag, 2003.

\bibitem{EdmondsIntersection}
J.~Edmonds.
\newblock Matroid intersection.
\newblock In P.~L. Hammer, E.~L. Johnson, and B.~H. Korte, editors, {\em
  Discrete Optimization I}, volume~4 of {\em Annals of Discrete Mathematics},
  pages 39--49. North-Holland, 1979.

\bibitem{Frank}
A.~Frank.
\newblock A weighted matroid intersection algorithm.
\newblock {\em Journal of Algorithms}, 2(4):328--336, 1981.

\bibitem{Fujishige}
S.~Fujishige.
\newblock {\em Submodular Functions and Optimization}, volume~58 of {\em Annals
  of Discrete Mathematics}.
\newblock Elsevier, second edition, 2005.

\bibitem{GabowXuFocs}
H.~N. Gabow and Y.~Xu.
\newblock Efficient algorithms for independent assignments on graphic and
  linear matroids.
\newblock In {\em Proceedings of the 30th Annual IEEE Symposium on Foundations
  of Computer Science (FOCS)}, pages 106--111, 1989.

\bibitem{GabowXu}
H.~N. Gabow and Y.~Xu.
\newblock Efficient theoretic and practical algorithms for linear matroid
  intersection problems.
\newblock {\em Journal of Computer and System Sciences}, 53(1):129--147, 1996.

\bibitem{Gantmacher}
F.~R. Gantmakher.
\newblock {\em The Theory of Matrices}, volume~1.
\newblock Chelsea, New York, 1960.
\newblock Translation by K.A. Kirsch.

\bibitem{GeelenThesis}
J.~F. Geelen.
\newblock {\em Matroids, Matchings, and Unimodular Matrices}.
\newblock PhD thesis, University of Waterloo, Canada, 1995.

\bibitem{GeelenCourse}
J.~F. Geelen.
\newblock Matching theory.
\newblock Lecture notes from the Euler Institute for Discrete Mathematics and
  its Applications, 2001.

\bibitem{LawlerAlg}
E.~L. Lawler.
\newblock Matroid intersection algorithms.
\newblock {\em Mathematical Programming}, 9:31--56, 1975.

\bibitem{RandomMatching}
L.~Lov\'asz.
\newblock On determinants, matchings and random algorithms.
\newblock In L.~Budach, editor, {\em Fundamentals of Computation Theory, FCT
  '79}, pages 565--574. Akademie-Verlag, Berlin, 1979.

\bibitem{LovaszPlummer}
L.~Lov\'asz and M.~D. Plummer.
\newblock {\em Matching Theory}.
\newblock Akad\'emiai Kiad\'o -- North Holland, Budapest, 1986.

\bibitem{Mucha}
M.~Mucha and P.~Sankowski.
\newblock Maximum matchings via {G}aussian elimination.
\newblock In {\em Proceedings of the 45th Annual IEEE Symposium on Foundations
  of Computer Science (FOCS)}, pages 248--255, 2004.

\bibitem{MurotaValuatedMatroid}
K.~Murota.
\newblock Valuated matroid intersection {II}: algorithms.
\newblock {\em SIAM Journal on Discrete Mathematics}, 9(4):562--576, 1971.

\bibitem{Murota}
K.~Murota.
\newblock {\em Matrices and Matroids for Systems Analysis}.
\newblock Springer-Verlag, 2000.

\bibitem{Pap}
G.~Pap.
\newblock A combinatorial algorithm to find a maximum even-factor.
\newblock In {\em Proceedings of the 11th International Conference on Integer
  Programming and Combinatorial Optimization (IPCO)}, pages 66--80, 2005.

\bibitem{RabinVazirani}
M.~O. Rabin and V.~V. Vazirani.
\newblock Maximum matchings in general graphs through randomization.
\newblock {\em Journal of Algorithms}, 10(4):557--567, 1989.

\bibitem{SankowskiParallel}
P.~Sankowski.
\newblock Processor efficient parallel matching.
\newblock In {\em Proceedings of the 17th ACM Symposium on Parallelism in
  Algorithms and Architectures (SPAA)}, pages 165--170, 2005.

\bibitem{Schrijver}
A.~Schrijver.
\newblock {\em Combinatorial Optimization: Polyhedra and Efficiency}.
\newblock Springer-Verlag, 2003.

\bibitem{Strang}
G.~Strang.
\newblock {\em Linear Algebra and its Applications}.
\newblock Thomson Learning, 1988.

\bibitem{Takazawa}
K.~Takazawa.
\newblock A weighted even factor algorithm.
\newblock Technical Report METR 2005-17, Department of Mathematical
  Informatics, University of Tokyo, July 2005.

\bibitem{Tomizawa}
N.~Tomizawa and M.~Iri.
\newblock {An Algorithm for Determining the Rank of a Triple Matrix Product
  {$AXB$} with Application to the Problem of Discerning the Existence of the
  Unique Solution in a Network}.
\newblock {\em Electronics and Communications in Japan (Scripta Electronica
  Japonica II)}, 57(11):50--57, Nov. 1974.

\end{thebibliography}
\bibliographystyle{abbrv}
\end{small}

%%%%%%%%%%%%%%%%%%%%%%%%%%%%%%%%%%%%%%%%%%%%%%%%%%%%%%%%%%%%%%%%%%%%%%%%%%%%%%%%%%%%%%%%%%%%%%%%%%%%

\end{document}